\documentclass[twocolumn,showpacs,preprintnumbers,amsmath,amssymb]{revtex4}
\usepackage{graphicx,psfrag}% Include figure files
\usepackage{dcolumn}% Align table columns on decimal point
\usepackage{bm}% bold math

\newcommand{\bea}{\begin{eqnarray}}
\newcommand{\eea}{\end{eqnarray}}
\newcommand{\simgt}{\hbox{ \raise3pt\hbox to 0pt{$>$}\raise-3pt\hbox{$\sim$} }}
\newcommand{\simlt}{\hbox{ \raise3pt\hbox to 0pt{$<$}\raise-3pt\hbox{$\sim$} }}

\begin{document}

\preprint{TU--884}

\title{Boost-invariant Leptonic Observables and Reconstruction of Parent Particle Mass}% Force line breaks with \\

\author{Sayaka~Kawabata$^a$, Yasuhiro~Shimizu$^{a,b}$, Yukinari~Sumino$^a$ and Hiroshi~Yokoya$^c$}
\affiliation{
$^a$Department of Physics, Tohoku University,
Sendai, 980--8578 Japan
\\
$^b$IIAIR, Tohoku University,
Sendai, 980--8578 Japan
\\
%$^c$Department of Physics and National Center for Theoretical Sciences, National Taiwan University, 
$^c$NCTS, National Taiwan University, 
Taipei 10617, Taiwan
}%

\date{\today}% It is always \today, today,

\begin{abstract}
We propose a class of observables constructed from lepton energy distribution, which are independent of
the velocity of the parent particle if it is scalar or unpolarized.
These observables may be used to measure properties of various particles in the LHC experiments.
We demonstrate their usage in a determination of the Higgs boson mass. 
%generated via the vector-boson fusion process.
\end{abstract}

\maketitle

The data and analyses from the experiments at the CERN Large Hadron Collider
(LHC) are attracting increasing attention.
The main goals of these experiments are discoveries of the Higgs boson 
and of signals of physics beyond the Standard Model.
Once the Higgs boson or other new particles are found, the next step is to uncover
properties of these particles.
There are well-known challenges in performing accurate measurements for such purposes:
(i) In reconstructing the kinematics of events, it is difficult to 
reconstruct jet energy scales accurately.
This is in contrast to electron and muon energy-momenta, which can be measured
fairly accurately.
(ii) Often interesting events include undetected particles which carry off missing momenta.
In this case, reconstruction of missing momenta is non-trivial.
(iii) The parton distribution function (PDF) of proton is needed in predicting cross sections and 
kinematical distributions.
Our current knowledge of PDF is limited, which induces 
relatively large uncertainties in these predictions.
As a consequence of these difficulties, for instance,
it is difficult to measure  
the energy-momentum of a produced new particle or predict accurately their
 statistical distributions.

To circumvent these difficulties, many sophisticated
methods for kinematical reconstruction of events have been devised.
See, for instance, \cite{Barr:2003rg,Barr:2010zj}
and references therein.
In these methods, one takes advantage of various
kinematical constraints, which follow from specific event topologies, 
to complement uncertainties induced by
the above difficulties.
Still, in most cases challenges remain to reduce systematic uncertainties
originating from ambiguities in jet energy scales, PDF and other non-perturbative
or higher-order QCD
effects. 
Generally it is quite non-trivial to keep these uncertainties 
under control, both theoretically and experimentally.

In this paper we propose a class of observables which can be used to
measure properties of particles produced in the LHC experiments, by largely
avoiding the above uncertainties.
We consider a particle $X$ which is scalar or unpolarized and whose decay
daughters include one or more charged leptons $\ell$$=e^\pm$ or $\mu^\pm$.
The observables are constructed from the energy distribution of $\ell$
and are independent of the velocity of $X$.
%These observables are independent of jet momenta.
% Only when computing small corrections in order
% to  incorporate the effects induced by 
% experimental cuts and backgrounds, jet momenta and information on the
% initial-state are involved.
% Therefore, systematic uncertainties are suppressed and 
% can be kept under control.
Although 
experimental cuts and backgrounds induce corrections to this
property,
we will show that systematic uncertainties are suppressed and 
can be kept under control.
Furthermore, we can utilize the degree of freedom of the 
observables to reduce or control effects induced by cuts
and backgrounds.

In the first part of the paper, we explain the construction of these observables,
in the two-body decay and multi-body decay cases separately.
In the latter part, we demonstrate its usefulness in a determination of
the Higgs boson mass using the vector-boson fusion process.

\medbreak
\noindent
{2-body decay: {$ X\to \ell + Y$}}

Suppose a parent particle $X$ decays into two particles, one of which
is a lepton
$\ell$ whose energy
is $ E_0$ (monochromatic) in the rest frame of $X$.
In the case that $X$ is scalar or unpolarized, the normalized lepton 
energy distribution in a boosted frame, in which $X$ has
a velocity  $\beta$, is given by
\bea
{\cal D}_\beta(E_\ell;E_0)=\frac{1}{(e^y-e^{-y})E_0}\,
\theta(e^{-y}E_0<E_\ell<e^y E_0) ,
\eea
in the limit where the mass of the lepton is neglected.
Here, $y$ is the rapidity of $X$ in the boost direction\footnote{
$y$ is defined with respect to the boost
direction of $X$.
This should be distinguished from the (pseudo-) rapidity
$\eta$, defined with respect to the beam direction
of experiments, used in the simulation study
of $m_H$ reconstruction.
},
 related to
$\beta$ as $e^{2y}=(1+\beta)/(1-\beta)$.
The step function is defined such that
$\theta(cond.)=1$ if $cond.$ is satisfied, and $\theta(cond.)=0$ otherwise.

We construct an observable $g(E_\ell/E_0)$ from the lepton energy $E_\ell$
in the boosted frame such that 
the expectation value $\langle g \rangle$
is independent of $\beta$.
For later convenience, we
write
$
g(E_\ell/E_0)=[dG(x)/dx]_{x=E_\ell/E_0} 
$.
Hence, 
\bea
\left\langle g\right\rangle \! = \!
\int \!\! dE_\ell\, {\cal D}_\beta (E_\ell;E_0) \,  g(E_\ell/E_0)
= \frac{G(e^y)-G(e^{-y})}{e^y-e^{-y}} \, .
\eea 
This shows that  $\langle g \rangle$ is the even part $F(y)+F(-y)$
of $F(y)\equiv G(e^y)/(e^y-e^{-y})$.
It is $\beta$($y$)-independent when $F$ is an odd function of $y$ plus a constant
independent of $y$.
Namely, $G(e^y)=(\mbox{even fn.\ of $y$})+const.\times(e^y-e^{-y})$.

Let us demonstrate usage of  $\langle g \rangle$.
In the case $G=e^y-e^{-y}$, we obtain $\langle 1/E_\ell^2 \rangle=1/E_0^2$.
Thus, (mathematically) we can reconstruct $E_0$ from the lepton energy distribution
irrespective of $\beta$ of the parent particle.
In the case $G(e^y)=(\mbox{even fn.\ of $y$})$,
$\left\langle g(E_\ell/E_0) \right\rangle =0$.
Conversely, we can adjust $E_0$, which enters $g(E_\ell/E_0)$ as an
external parameter,
such that $\langle g \rangle$ vanishes and determine
the true value of $E_0$.
We give two examples of $G$ for the latter case:
(a)
$G^{(a)}_n={1}/[2\,n\,{\cosh (ny)}]$, corresponding to
$g= E_0^{n+1}E_\ell^{n-1} (E_0^{2n}-E_\ell^{2n})/(E_0^{2n}+E_\ell^{2n})^2$.
As $n$ increases, contributions of the $E_\ell$ distribution from large $|y|$
region
($E_\ell \ll E_0$ or  $E_\ell\gg E_0$)
become more suppressed.
(b)
$G^{(b)}_r=\theta( r< e^{-|y|} < 1)(e^{-|y|}-r)$,
corresponding to 
$g = \theta( r < E_\ell/E_0 < 1) - (E_0/E_\ell)^2
\theta(1<E_\ell/E_0<r^{-1})$.
This observable is independent of the $E_\ell$ distribution
in the regions $E_\ell < r E_0$ and $E_\ell>r^{-1}E_0$.
\medbreak

\noindent
{Many-body decay: {$ X\to \ell + {\rm anything}$}}

We assume that we know the theoretical prediction for the lepton energy
distribution in the rest frame of the parent particle $X$,
$
{d\Gamma_{X\to\ell+{\rm anything}}}/{dE_0}\,,
$
where $E_0$ is the lepton energy in this frame.
We define an observable in the boosted frame as
\bea
{\cal O}_G\equiv {\cal N} \int  \frac{dE_0}{E_0^2}\, 
\frac{d\Gamma_{X\to\ell+{\rm anything}}}{dE_0}\,
\left[\frac{d}{dx}\,G(x)\right]_{x=E_\ell/E_0} \,.
\label{defOG}
\eea
It depends on the lepton energy $E_\ell$ in the boosted frame
and on the parameters of $d\Gamma/dE_0$
such as the parent particle mass $m_X$.
$\cal N$ is an arbitrary normalization constant independent of $E_l$.
We can prove (see below) that, if $X$ is scalar or unpolarized, and
if we take the same $G$ as in the 2-body decay case,
$\langle {\cal O}_G \rangle$ is independent of $\beta$ of $X$.
In particular, in the case that $G(e^y)$ is an even function of $y$,
$\langle {\cal O}_G \rangle=0$.
The parent particle mass enters as an external parameter in the definition
of $ {\cal O}_G$.
Hence, we can use this property to determine $m_X$, provided that
other parameters are known.
$g$ in the 2-body decay case can be regarded as a special
case of $ {\cal O}_G$.

{\it Proof:}~
The lepton energy distribution in the boosted frame is given by
\bea
f_\beta (E_\ell)=\int dE'_0 \, \frac{d\Gamma_{X\to\ell+{\rm anything}}}{dE'_0}\,
%f_\beta (E_\ell)=\int {dE'_0} \, \frac{d\Gamma_{X\to\ell+{\rm anything}}}{dE'_0}\,
{{\cal D}_\beta (E_\ell;E'_0)}
%{\Gamma_{X\to\ell+{\rm anything}}} 
\, .
\eea
Hence,
\bea
&&
\langle {\cal O}_G \rangle=\int dE_\ell \, f_\beta(E_\ell)\,  {\cal O}_G
\nonumber\\&&
= {\cal N} \int dE_0dE'_0\, \frac{d\Gamma}{dE_0}\frac{d\Gamma}{dE'_0}\,
\frac{G(e^y E'_0/E_0) - G(e^{-y} E'_0/E_0)}{E_0\,E'_0\,(e^y-e^{-y})} \,,
\nonumber\\
\eea
where we integrated over $E_\ell$.
In the case $G(x^{-1})=G(x)$, 
$G(e^y E'_0/E_0) - G(e^{-y} E'_0/E_0)=G(e^y E'_0/E_0) - G(e^{y} E_0/E'_0)$
is anti-symmetric under the exchange of $E_0$ and $E'_0$, while all other
parts are symmetric.
It follows that $\langle {\cal O}_G \rangle=0$.
In the case $G(x)=x-x^{-1}$,
$G(e^y E'_0/E_0) - G(e^{-y} E'_0/E_0)=
(E'_0/E_0+E_0/E'_0)(e^y-e^{-y})$, so that 
$y$-dependence cancels out. (Q.E.D.)
\medbreak

Since we must know the theoretical prediction for $d\Gamma/dE_0$
to construct ${\cal O}_G $, we consider
use of ${\cal O}_G $ mainly for the purpose of
precision measurements, after
the nature of the parent particle (such as its decay modes)
is roughly determined by other 
methods.
In principle, different functions $G$ in constructing ${\cal O}_G$ 
may be used to determine simultaneously
more than one parameters in a decay process.
In this first study, however, we consider the case where only one
parameter is unknown.

Under a realistic experimental condition, in which $\beta$ of
$X$ has a distribution $D_X(\beta)$,
the lepton energy distribution in the laboratory frame
is given by
$
D(E_\ell)=\int d\beta\, D_X(\beta)f_\beta (E_\ell) 
$.
In this case, the expectation value
$\langle {\cal O}_G \rangle_D \equiv
\int dE_\ell \, D(E_\ell)\,  {\cal O}_G$ 
is independent of $D_X$.
Hereafter, we choose $G$ to be an even function of $y$ and
use $\langle {\cal O}_G \rangle_D=0$ to determine $m_X$.
Generally various experimental cuts, which
affect $D(E_\ell)$, are imposed
due to detector
acceptance effects and for event selection purposes.
Furthermore, there are contributions from background processes 
which also modify $D(E_\ell)$.
After incorporating these effects, there is no guarantee for 
$\langle {\cal O}_G \rangle$ to vanish.
Suppose that $D(E_\ell)$ is modified to $D(E_\ell)+\delta D(E_\ell)$
by these effects, where the distributions are normalized
as $\int dE_\ell \, D=\int dE_\ell \, (D+\delta D)
=1$.
In the case $|\delta D/D| \ll 1$,
if 
%instead of $\langle {\cal O}_G \rangle_D=0$,
$\langle {\cal O}_G \rangle_{D+\delta D}=0$ is used to extract $m_X$, 
the obtained value is systematically shifted from the true value 
by an amount
\bea
\delta m_X= -  \langle {\cal O}_G \rangle_{\delta D} \biggl/
\left\langle \frac{\partial {\cal O}_G}{\partial m_X} \right\rangle_D
,
\label{deltamX}
\eea
where we neglected ${ O}(\delta D^2)$ corrections.
We can use this formula to study systematic corrections 
%that need to be incorporated
in the determination of $m_X$.
$\delta D$ should be estimated by Monte Carlo (MC) simulations which 
take into account realistic
experimental conditions.
Errors in the estimate of $\delta D$ contribute as systematic
uncertainties in the determination of $m_X$.\footnote{
We note that if the shape of the lepton energy distribution
is unchanged, i.e.\
$\delta D\propto D$, $\delta D$
does not affect the reconstructed mass value.
}

The production of Higgs bosons via vector-boson fusions is
expected to be observed with a good signal-to-noise ratio,
if the Higgs boson
mass $m_H$ is within the range $135~{\rm GeV}\simlt m_H \simlt 190~{\rm GeV}$,
hence it is considered as a promising channel for the Higgs boson
discovery. 
We perform a MC simulation to study feasibility of $m_H$
determination using the observable ${\cal O}_G$ and the
decay modes $H\to WW^{(*)}\to \ell\ell\nu\nu$
($\ell\ell=e\mu, \mu\mu, ee$).
We generate the events for the signal and background processes
using MadEvent \cite{Maltoni:2002qb}, which are passed to PYTHIA 
\cite{Sjostrand:2006za} and then to the
fast detector simulator PGS \cite{PGS}.
We set $\sqrt{s}=14$~TeV.
% and use ... for PDF.

The strategy of our analysis follows, to a large extent, that of
\cite{Asai:2004ws}, which studied the prospect of Higgs boson search using the
vector-boson fusion process in the ATLAS experiment.
We first repeated the analysis of \cite{Asai:2004ws} in the case of
$H\to e\mu\nu\nu$ mode %($m_H=160$~GeV)
using our analysis tools and imposing the same cuts.
We reproduced the numbers in Tab.~4 of \cite{Asai:2004ws}
reasonably well, considering differences
in both analyses, such as different
detector simulators, different PDF's, and
different jet clustering algorithms
(we use the cone algorithm with $R=0.5$; 
we do not correct the jet energy scales given
by the output of PGS):
regarding the signal events,
we reproduced
the efficiencies of the cuts involving only leptons within 
a few~\% accuracy and those involving jets within
a few tens~\% accuracy; in total, the efficiency of all the cuts was
reproduced with 6\% accuracy.
On the other hand, the cross section of the signal 
%($t\bar{t}+tW$ background)
is smaller by 30\% 
%(larger by 25\%) 
in our analysis as compared to
that of \cite{Asai:2004ws}.
The difference originates from the different 
scales and different
PDF's used in the event generators.
We do not correct the difference in normalization
of the cross sections;
it may result in overestimates of the statistical errors given below.
%hence our error estimates given below may be
%conservative.

For simplicity, in our analysis
we omit the Higgs boson production via $gg$ fusion and the
Higgs decay modes including $\tau$'s.
In principle, part of these modes including $e$ or $\mu$ in the final states 
can be used as 
signal events, since $d\Gamma_{H\to\ell+{\rm anything}}/dE_0$
is calculable.

In our analysis of $m_H$ reconstruction, ideally
two criteria need to be satisfied to ensure use of the observable 
${\cal O}_G$:
(1)~The lepton energy distribution in the Higgs rest frame agrees 
with the theoretical prediction
$d\Gamma_{H\to \ell\ell\nu\nu}/dE_0$.
(2)~The lepton angular distribution in the Higgs rest frame
is isotropic.
The effects of cuts should not
violate these criteria significantly.

Jets in the signal process are associated with the Higgs production
process and are independent of the Higgs decay process.
Therefore, cuts involving only jets would not affect the above
criteria but only affect the $\beta$ distribution of the Higgs boson.
By contrast, cuts involving leptons can affect the
above criteria significantly.

Taking this into account, we use the same
cuts as in \cite{Asai:2004ws} except the following two
cuts:\vspace*{2mm}\\
\noindent
(A)\,Lepton acceptance\\
~~~~~~~No muon isolation requirement;\\
~~~~~~~$P_T(e)>15$~GeV, $P_T(\mu)>10$~GeV,~$|\eta_\ell|<2.5$;\\
~~~~~~~Only one pair of identified leptons in each event.\\
(B)\,Lepton cuts\\ 
~~~~~~~$M_{\ell\ell}<45$~GeV,~~~$P_T(e),P_T(\mu)<120$~GeV.
\vspace*{2mm}
\\
Concerning
(A), since  lepton isolation requirement as well as cuts on lepton $P_T$ and $\eta$
bias the lepton angular distribution in the Higgs rest frame, 
we loosen the cuts and requirement as much as possible.
Concerning (B), it is important to select events with
leptons in the same directions,
in order to reduce background events.
% Ref.~\cite{Asai:2004ws} imposed a combination of cuts on the
% angles of leptons in the laboratory frame and a cut on $M_{\ell\ell}$.
We note that, if we impose the same cut on $M_{\ell\ell}$ in the
theoretical prediction for $d\Gamma_{H\to \ell\ell\nu\nu}/dE_0$,
this cut does not contribute to systematic errors.
Hence, we tighten the Lorentz invariant cut and omit other frame-dependent 
angular cuts.
By modification of the cuts (A) and (B), the efficiency of the signal events reduces by
a few tens~\%, while variations of
the efficiencies of the background events are small.
\footnote{
$m_H$-dependent cut on the transverse mass $M_T$ of $\ell\ell$-$P_T^{miss}$ system
(used in Tab.~7 of \cite{Asai:2004ws}) is omitted.
We expect
many additional ways for optimization 
in the $m_H$ determination, such as inclusion of this cut.
}

\begin{table}[t]
\begin{center}
\begin{ruledtabular}
\footnotesize
\begin{tabular}{l|ccc|cc}
%\hline
&\multicolumn{3}{c|}{Signal ($m_H$)}&\multicolumn{2}{c}{Background}\\
&150~GeV&180~GeV&200~GeV&$t\bar{t}+Wt$&$WW$+jets\\
\hline
$e\mu$ mode\,[fb]&2.00&2.21&1.07&0.53&0.13\\
$\mu\mu$ mode\,[fb]&1.33&1.47&0.68&0.31&0.05\\
$ee$ mode\,[fb]&0.76&0.89&0.42&0.24&0.04\\
% \hline
% $\Delta m_H^{\rm stat.}$[GeV]& $4.1\sqrt{\frac{818}{N_\ell}}$&
% $10\sqrt{\frac{914}{N_\ell}}$&$14\sqrt{\frac{434}{N_\ell}}$&&\\
%\hline
\end{tabular}
\vspace*{-2mm}
\end{ruledtabular}
\caption{\small
Cross section$\times$efficiency, after all the cuts.
$e,\mu$ from $\tau$ are not included.
\label{tabcs-eff}
}
\end{center}
%\vspace*{-5mm}
\end{table}

Tab.\,\ref{tabcs-eff} lists estimates of 
(cross section)$\times$(efficiency) 
of the signal and background events,
after all the cuts are imposed.%
\footnote{After this paper was submitted to arxiv, the ATLAS and CMS collaborations announced new Higgs mass bounds at the Lepton-Photon 2011 Conference
\cite{LP2011}.
Although the parameters $m_H=150$, $180$, $200$ GeV are excluded at 95~\% C.L.,
the results are still preliminary and the exclusions are not conclusive. In particular,  $m_H=150$~GeV is at the
boundary of the 95~\% C.L. excluded region and it would be premature to exclude this case.}
For the backgrounds, we simulate only $t\bar{t}+Wt$
and $WW$+jet (electroweak) events, which are shown to be
the major backgrounds in \cite{Asai:2004ws}.
% Other background events are less than 2\% of the
% signal events, and w
We estimate contributions of other backgrounds to be negligible,
compared to the uncertainties discussed below.
Consistently with the signal, we omit background events
with $\ell=e,\mu$ from $\tau$.

% The bottom line of Tab.~I shows estimates of
% the statistical uncertainties %(standard deviations) 
% in the $m_H$ determination, in the case
% that $N_\ell$ leptons are used to take the expectation value
% $\langle {\cal O}_G \rangle$.
% The factors in the square-root correspond to unity for
% an integrated luminosity of 100\,fb$^{-1}$ and collecting all
% $\ell$ from the three modes $\ell\ell=e\mu$, $\mu\mu$, $ee$ of
% the signal events.
% These estimates are derived in the following way.
% We generate large-statistics MC events and separate them into
% ensembles. 
% We perform a pseudo-experiment to reconstruct $m_H$ for each ensemble
% and examine the statistical distribution of reconstructed
% $m_H$.
% After checking that the distribution is close to a Gaussian distribution,
% the standard deviation of the distribution is assigned as an estimate of
% the statistical error.\footnote{
% In order to increase the number of usable events, we did not apply the cuts,
% which involve only jets, in the estimates of statistical errors, 
% assuming that effects of these cuts are minor.
% }
% We use ${\cal O}_G$ with $G=G^{(a)}_{n=4}$ in these estimates
% (see below).

We test the two criteria with the 
signal events which passed all the
cuts.
The histogram in
Fig.~1 shows the lepton ($\ell = \mu$) energy distribution in
the Higgs rest frame for the $H\to \mu\mu\nu\nu$ mode
and MC input $m_H^{\rm MC}=150$~GeV, 
where we looked up the parton-level neutrino
momenta in each event to reconstruct the Higgs momentum.
We generated a large-statistics event sample 
(about 12,000 events after applying all the cuts) in order to focus on
the systematic effects caused by the cuts.
The theoretical prediction for $d\Gamma_{H\to \ell\ell\nu\nu}/dE_0$
with the cut $M_{\ell\ell}<45$~GeV 
(no other cuts are applied) is also plotted
with a red line.
A good agreement is observed.
%, even though the efficiency of all the
%cuts is only about 7\%.

\begin{figure}[t]%\centering
\includegraphics[width=6cm]{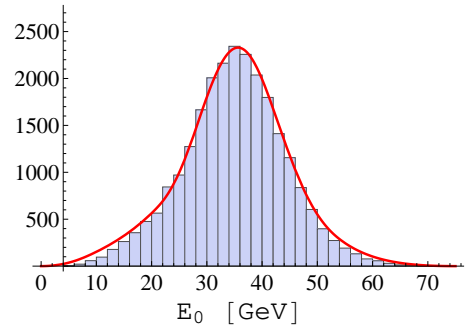}
\vspace*{-2mm}
\caption{\small
Lepton energy distribution in the Higgs rest frame.
\label{Fig1}
}
\vspace{-3mm}
\end{figure}

Figs.~2 show the lepton $\cos\theta$ distributions in the Higgs rest 
frame of these events, 
where $\theta$ is the angle measured from the Higgs boost direction.
The total event sample is divided into four 
groups of equal size,
in the increasing order of $\beta$ (or $y$) of the Higgs boson.
The average $y$ value for each group is displayed.
The events with low $y$ have a distribution closer to
isotropic one, whereas
the events with large boost factors have
strong distortion of the $\cos\theta$ distribution.
Since Higgs bosons with large $y$ are boosted mostly 
in the beam directions,
the lepton $P_T$ and $\eta$ cuts and electron isolation requirement bias
the $\cos\theta$ distribution.
In particular, depletion of events in the $\cos\theta\simeq 1$ region 
of large $y$ samples is
caused by the lepton $\eta$ cuts.
On the other hand,
Higgs bosons with small $y$ are boosted
in random directions, so that the lepton acceptance effects do
not bias the lepton $\cos\theta$ distribution strongly.
Thus, the second criterion is satisfied only by
events with small boost factors.

We may choose $G$ appropriately to suppress contributions of
events with large $y$.
By examining various $G$, we find that
$G=G^{(a)}_n$ for $n\approx 4$ is an optimal choice.
As $n$ increases, $\langle {\cal O}_G \rangle$ becomes
less sensitive to events with large $y$ but more sensitive
to statistical fluctuations.\footnote{
The statistical error of reconstructed
$m_H$ increases gradually with $n$.
% If good estimates of 
% $\delta m_H$ are achieved, one might as well decrease
% $n$ to reduce the statistical error.
}
% We also tested $G^{(b)}_r$ for different $r$.
% The performance of an optimal observable is similar to that with
% $G^{(a)}_{n=4}$.

\begin{figure}[t]%\centering
\begin{tabular}{cc}
\includegraphics[width=3.5cm]{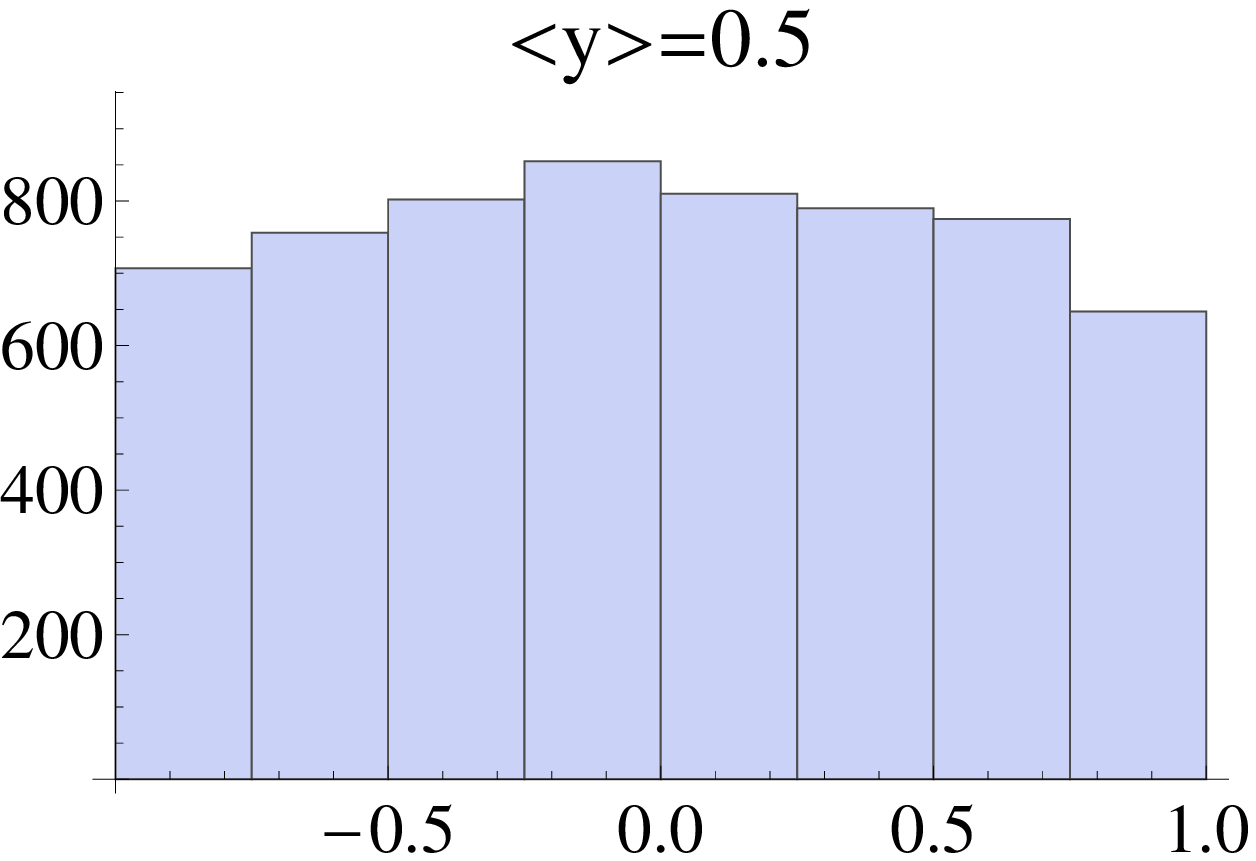}&
\includegraphics[width=3.5cm]{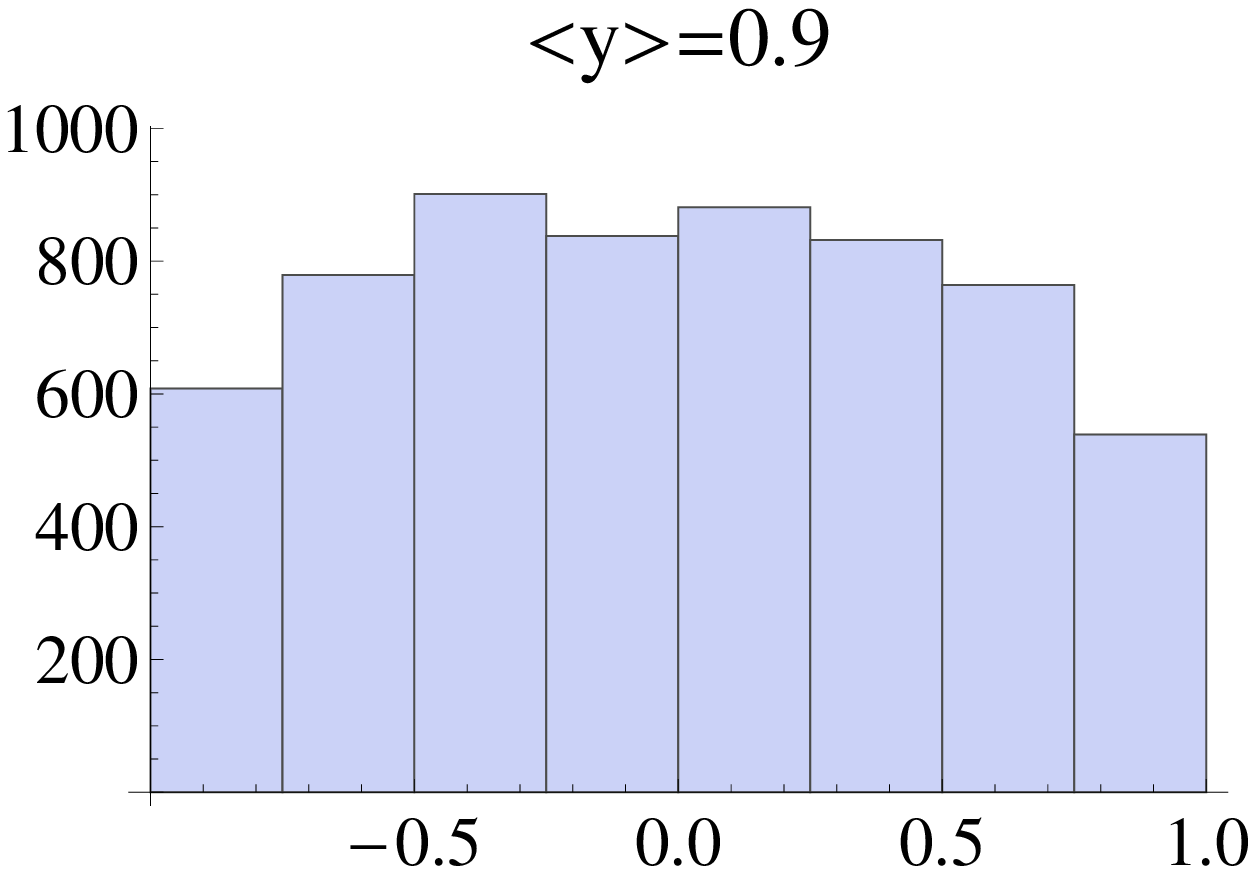}\vspace*{2mm}\\
\includegraphics[width=3.5cm]{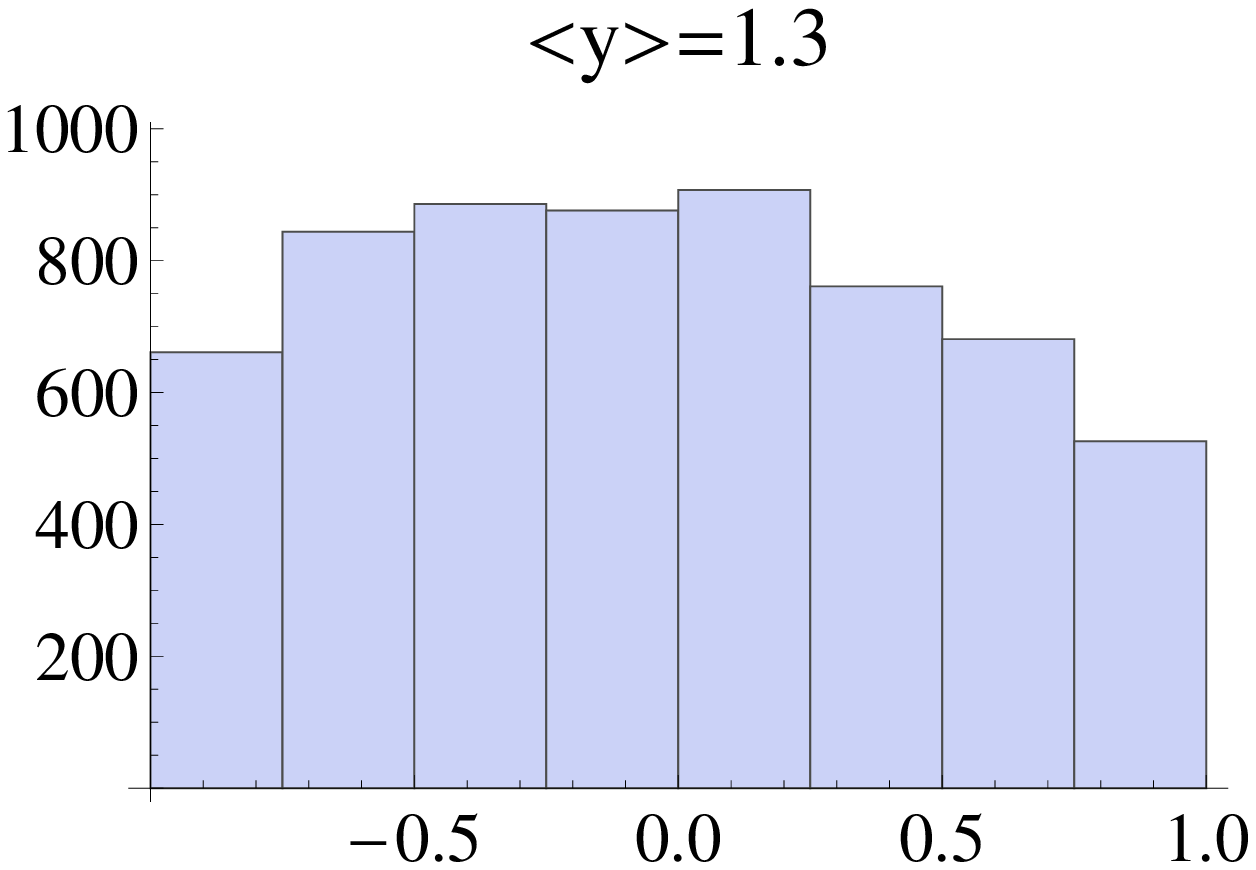}&
\includegraphics[width=3.5cm]{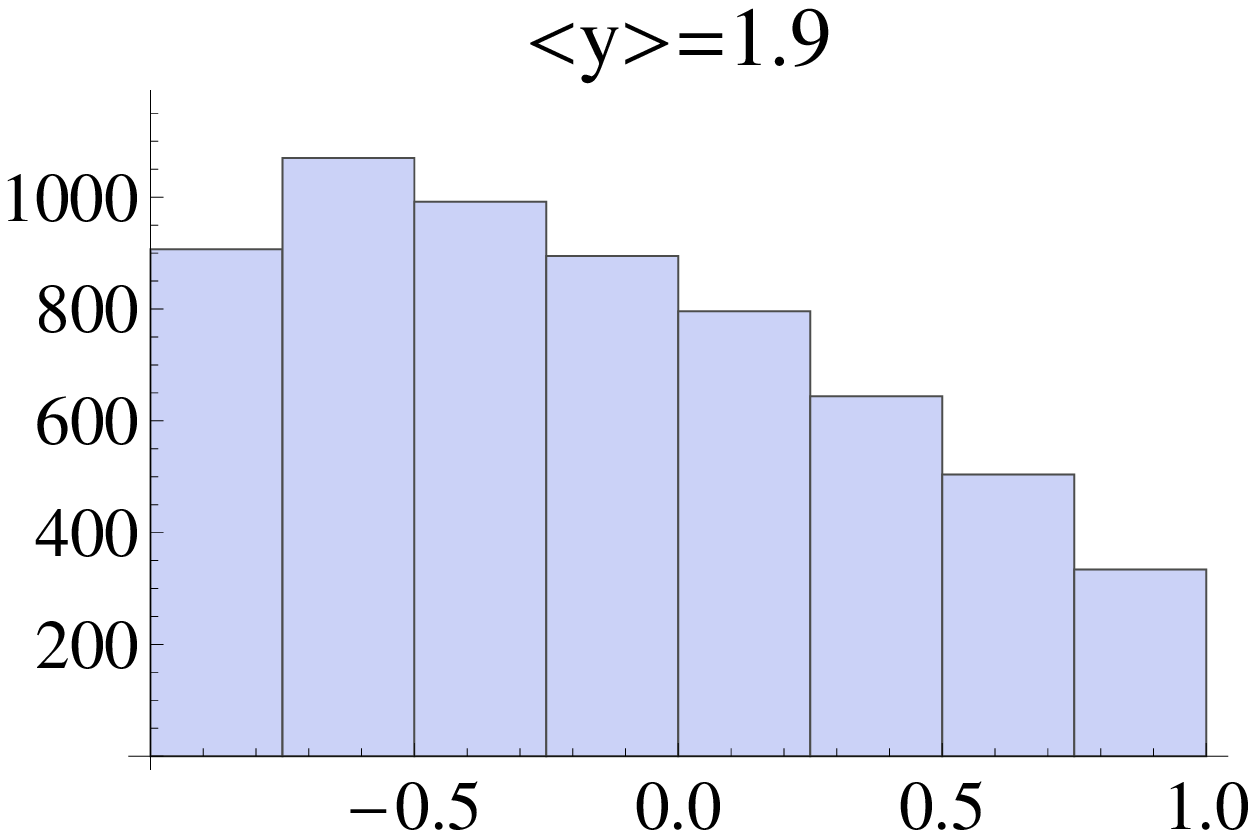}\\
\end{tabular}
\vspace*{-2mm}
\caption{\small
\label{Fig2} Lepton $\cos\theta$ distribution in the Higgs rest 
frame.
}
\end{figure}

\begin{figure}[t]%\centering
\includegraphics[width=6cm]{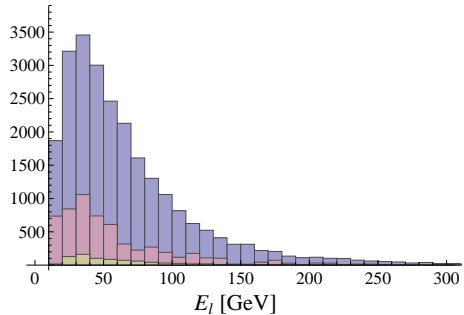}
%\vspace*{-2mm}
\caption{\small
Muon energy distribution
after applying all the cuts  for $H\to \mu\mu\nu\nu$ mode.
Histograms represent muons from
the signal (corresponding to $m_H^{\rm MC}=150$~GeV),
$t\bar{t}+Wt$ and $WW$+jet events (overlayed, from back to front).
\label{Fig3}
}
\vspace{-3mm}
\end{figure}

\begin{figure}[t]%\centering
\includegraphics[width=7cm]{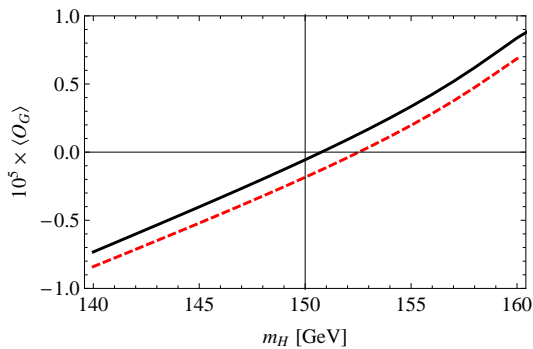}
%\vspace*{-2mm}
\caption{\small
$10^5\times\langle {\cal O}_G \rangle$ as a function of 
$m_H$ in the theoretical prediction for
$d\Gamma_{H\to \ell\ell\nu\nu}/dE_0|_{M_{\ell\ell}<45{\rm GeV}}$.
Muon energy distribution given in Fig.~\ref{Fig3} is used:
solid (dashed) line represents the expectation value taken with
respect to the signal--plus--background (signal) MC events.
\label{Fig4}
}
%\vspace{-3mm}
\end{figure}

In $m_H$ determination, typically
there are two solutions which satisfy $\langle {\cal O}_G \rangle=0$,
one above and one below the $WW$ threshold $2M_W$.
We believe that, unless $m_H$ is very close to $2M_W$ \footnote{
This mass region is being excluded by the recent Tevatron searches
\cite{Aaltonen:2011gs}.
}, the correct
solution can be identified relatively easily, using profiles of
the lepton energy distribution, dilepton invariant mass distribution,
etc.
Hence, we consider deviations only around the correct
solution.

Fig.~\ref{Fig3} shows the muon energy distributions
separately for the signal events and 
the $t\bar{t}+Wt$ and $WW$+jet background events,
which passed through all the cuts for the $H\to \mu\mu\nu\nu$ 
mode.
The signal events are generated with $m_H^{\rm MC}=150$~GeV 
and the number of events after the cuts is
about 12,000.
The normalizations of the background events 
have been rescaled according to the respective cross sections to 
match the number of the signal events.
Using this distribution the expectation value
$\langle {\cal O}_G \rangle$ 
is computed;
see Fig.~\ref{Fig4}.\footnote{
In Figs.~\ref{Fig4} and \ref{Fig-VaryCuts}, we choose the
normalization of ${\cal O}_G$ as 
$
{\cal N}= 1/\Gamma(H\to{\mu\mu\nu\nu};{M_{\mu\mu}<45\,{\rm GeV}})
\,.
$
}
The observable
${\cal O}_G$ in eq.~(\ref{defOG})
is defined with $G=G^{(a)}_{n=4}$ and
the theoretical prediction for 
$d\Gamma_{H\to \ell\ell\nu\nu}/dE_0$
is computed imposing
a cut ${M_{\ell\ell}<45~{\rm GeV}}$ on the dilepton invariant mass.
The value of $\langle {\cal O}_G \rangle$ changes as a function
of $m_H$, which enters as an external parameter in the
theoretical prediction for
$d\Gamma_{H\to \ell\ell\nu\nu}/dE_0|_{M_{\ell\ell}<45{\rm GeV}}$.
Without any cuts and without background contributions,
$\langle {\cal O}_G \rangle$ would cross zero at 
$m_H=150$~GeV in the large statistics limit.\footnote{
The statistical error corresponding to the current
MC events ($N_\ell\approx 24,000$) is $\Delta m_H^{\rm stat.}\simeq 0.6$~GeV;
c.f.\ Tab.~\ref{tabstat-err}.
}
As seen in Fig.~\ref{Fig4}, due to the effects of the cuts, 
the expectation value $\langle {\cal O}_G \rangle$ taken
with respect to the signal events crosses zero
at about $+2.6$~GeV above the MC input value.
This value is consistent with $\delta m_H$ determined
from eq.~(\ref{deltamX}).
$\langle {\cal O}_G \rangle$ with respect to the 
signal--plus--background events crosses zero 
at $\delta m_H=+0.8$~GeV above the MC input value.\footnote{
In this particular example, however,
since $\delta m_H^{\rm sig+bkg}\sim\Delta m_H^{\rm stat.}$, we can
only estimate $\delta m_H^{\rm sig+bkg}\sim 0.8\pm 0.6$~GeV.
The present statistics of the MC events should be sufficient
for most of the other estimates of $\delta m_H$ in Tab.\,II.
}

\begin{table}[t] 
\begin{center}
\begin{ruledtabular}
\footnotesize
\begin{tabular}{l|lccc}
$\ell\ell$&
&$m_H=$150~GeV&$m_H=$180~GeV&$m_H=$200~GeV\\
\hline
& Signal & $+5.7$ & $-0.9$ & $-5.5$ \\
$e\mu$ & Bkg: $t\bar{t}+Wt$& $-4$ & $+14$ & $+16$ \\
& ~~~~~~$WW$+jets & $-1$ & $+3$ & $+5$ \\
\hline
& Signal & $+2.6$ & $+0.5$ & $-3.4$ \\
$\mu\mu$ & Bkg: $t\bar{t}+Wt$& $-2$ & $+11$ & $+16$ \\
& ~~~~~~$WW$+jets & $0$ & $0$ & $-1$ \\
\hline
& Signal & $+7.6$ & $-1.8$ & $-13.2$ \\
$ee$ & Bkg: $t\bar{t}+Wt$& $-1$ & $+5$ & $+5$ \\
& ~~~~~~$WW$+jets & $0$ & $+2$ & $+2$ \\
\end{tabular}
\vspace*{-2mm}
\end{ruledtabular}
\caption{\small
Estimate of systematic shift $\delta m_H$~[GeV],
as defined in eq.~(\ref{deltamX}).
(Corresponding systematic error is
different; see text.)
\label{delmH}
}
\end{center}
%\vspace{-3mm}
\end{table}

\begin{table}[t]
\begin{center}
\footnotesize
\begin{tabular}{l|c|c|c}
\hline
$m_H$ [GeV]&150&180&200\\
\hline
$e\mu$ mode [GeV]
&$4.1\sqrt{\frac{400}{N_\ell}}$&$11\sqrt{\frac{442}{N_\ell}}$&$14\sqrt{\frac{214}{N_\ell}}$\\
$\mu\mu$ mode [GeV]
&$5.5\sqrt{\frac{266}{N_\ell}}$&$14\sqrt{\frac{294}{N_\ell}}$&$20\sqrt{\frac{136}{N_\ell}}$\\
$ee$ mode [GeV]
&$5.7\sqrt{\frac{152}{N_\ell}}$&$14\sqrt{\frac{178}{N_\ell}}$&$18\sqrt{\frac{84}{N_\ell}}$\\
\hline
Combined [GeV]&$< 3.1\sqrt{\frac{818}{N_\ell^{\rm tot}}}$&
$< 8.2\sqrt{\frac{914}{N_\ell^{\rm tot}}}$&$< 11\sqrt{\frac{434}{N_\ell^{\rm tot}}}$\\
\hline
\end{tabular}
\caption{\small
Estimates of the statistical error $\Delta m_H^{\rm stat.}$ in the
$m_H$ reconstruction using $N_\ell$ leptons.
($e,\mu$ from $\tau$ are not included.)
The factors in the square-root correspond to unity for
an integrated luminosity of 100\,fb$^{-1}$ and using
leptons of the signal events.
($N_\ell^{\rm tot}=N_\ell^{(\mu\mu)}+N_\ell^{(e\mu)}+N_\ell^{(ee)}$.)
\label{tabstat-err}
}
\end{center}
%\vspace*{-5mm}
\end{table}

In a similar manner, we can estimate
the systematic shift 
$\delta m_H$, as defined in eq.~(\ref{deltamX}), in the
$m_H$ reconstruction by varying the input Higgs mass value
in MC simulations and using different modes.
The values of $\delta m_H$ for some sample points are listed in Tab.\,\ref{delmH}.
The magnitude of $\delta m_H$ due to all the cuts on the signal events
is smaller for $\ell\ell=\mu\mu$
than for $e\mu$ and is the largest for $ee$.
This is because the acceptance corrections are 
smaller for $\mu$ than $e$.
% , and due to a better 
% energy resolution for $\mu$ than $e$.
We confirm $|\delta m_H/m_H|\ll 1$, which shows that deviations from
the ideal limit is suppressed and kept under control.

% Since all the cuts are independent of the 
% assumed or reconstructed Higgs mass
% value,
% we can simply reconstruct $m_H$ from the relation $\langle {\cal O}_G \rangle=0$
% and correct it by the value of $\delta m_H$ listed in Tab.\,\ref{delmH}
% (or that estimated in a similar manner in case of a different $m_H$ value).
% This procedure works as long as $|\delta m_H/m_H| \ll 1$ and
% $\langle {\cal O}_G \rangle$ can be approximated by a linear function of $m_H$
% in the vicinity of the solution.
% As demonstrated above, $\delta m_H$
% can be estimated by estimating 
% $\delta D$ 
% (modification of the lepton energy distribution)
% using MC simulations.
% For backgrounds, $\delta D$ may also be measurable
%  experimentally.
% Hence, systematic errors for
% the final estimate of $m_H$
% are expected to be considerably smaller than the magnitudes
% of $\delta m_H$ in Tab.~\ref{delmH}.
% For instance, if we can estimate $\delta D$ originating
% from the $t\bar{t}+Wt$ background with 10\% accuracy,
% the systematic error is 10\% of the corresponding
% value in the table, which is below 2~GeV.

Tab.~\ref{tabstat-err} shows estimates of
the statistical uncertainties (standard deviations) 
in the $m_H$ determination, $\Delta m_H^{\rm stat.}$, in the case
that $N_\ell$ leptons are used to take the expectation value
$\langle {\cal O}_G \rangle$.
% \footnote{
% We used a large-statistics MC event sample for these estimates,
% assuming that effects of the cuts involving only jets are 
% negligible.
% }
The factors in the square-root correspond to unity for
an integrated luminosity of 100\,fb$^{-1}$ and using
leptons of the signal events in the respective modes.
The bottom row lists the upper bounds of combined statistical
errors of the three modes;
these are the largest among the three modes, when the
number of leptons is set as the sum of all three modes,
$N_\ell^{\rm tot}=N_\ell^{(\mu\mu)}+N_\ell^{(e\mu)}+N_\ell^{(ee)}$.
We use ${\cal O}_G$ with $G=G^{(a)}_{n=4}$ in these estimates.
The estimates are derived in the following way.
We compute $\delta{\cal O}_G^2\equiv \langle
{\cal O}_G^2 \rangle - \langle {\cal O}_G \rangle^2$
and $\partial \langle {\cal O}_G \rangle/\partial m_H$
using the MC signal events which passed all the cuts.
Since the statistical error of $\langle {\cal O}_G \rangle$
is given by $\Delta {\cal O}_G^{\rm stat.}=\delta{\cal O}_G/\sqrt{N_\ell}$,
we convert it to $\Delta m_H^{\rm stat.}$
using the tangent $\partial \langle {\cal O}_G \rangle/\partial m_H$
evaluated at each input value
$m_H=m_H^{\rm MC}$ ; c.f.\ Fig.~\ref{Fig4}.
% ensembles. 
% We perform a pseudo-experiment to reconstruct $m_H$ for each ensemble
% and examine the statistical distribution of reconstructed
% $m_H$.
% After checking that the distribution is close to a Gaussian distribution,
% the standard deviation of the distribution is assigned as an estimate of
% the statistical error.\footnote{
% In order to increase the number of usable events, we did not apply the cuts,
% which involve only jets, in the estimates of statistical errors, 
% assuming that effects of these cuts are minor.

\begin{widetext}
~
\begin{figure}[!]
\begin{tabular}{ll}
\includegraphics[width=8cm]{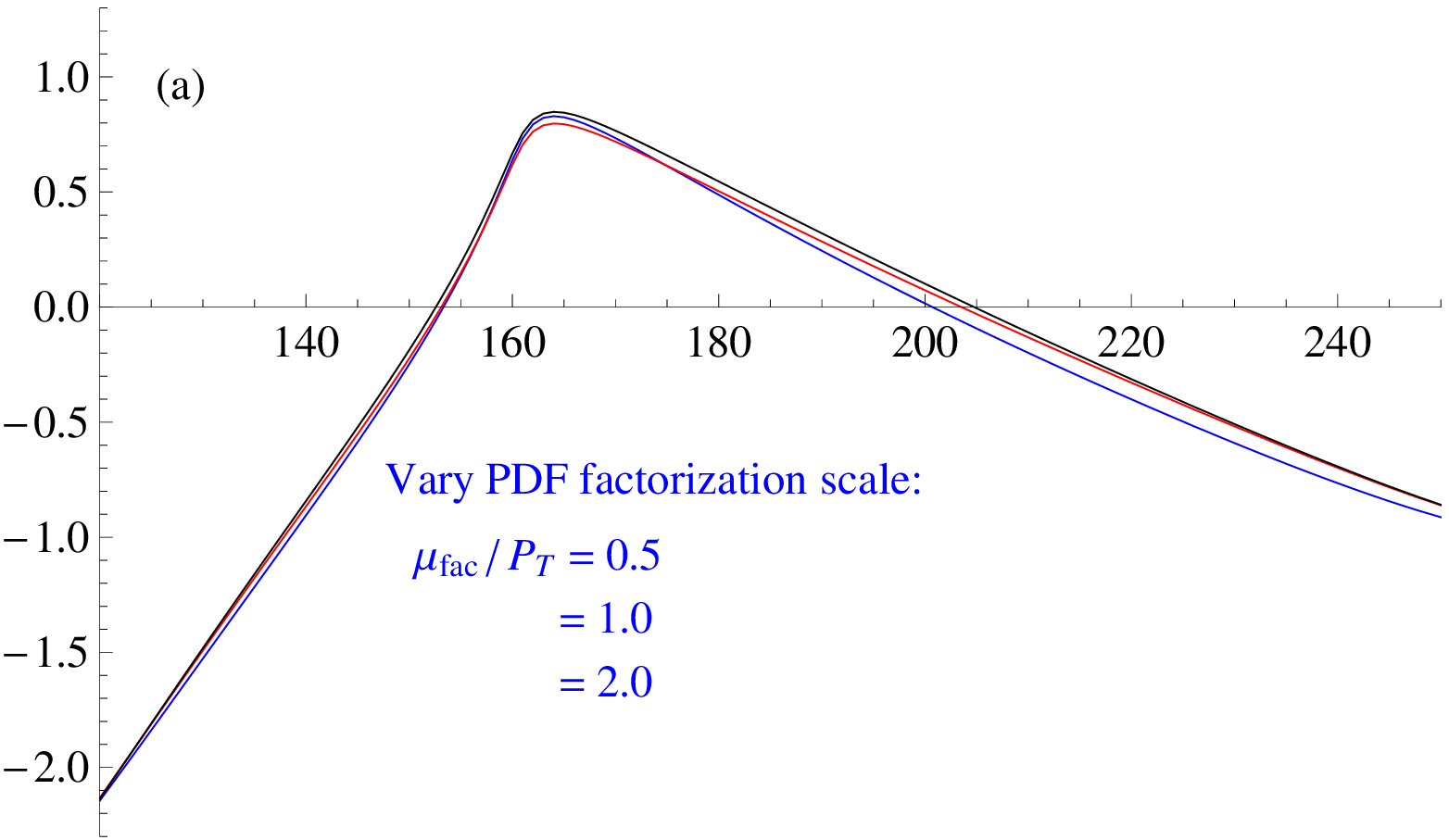}
\includegraphics[width=8cm]{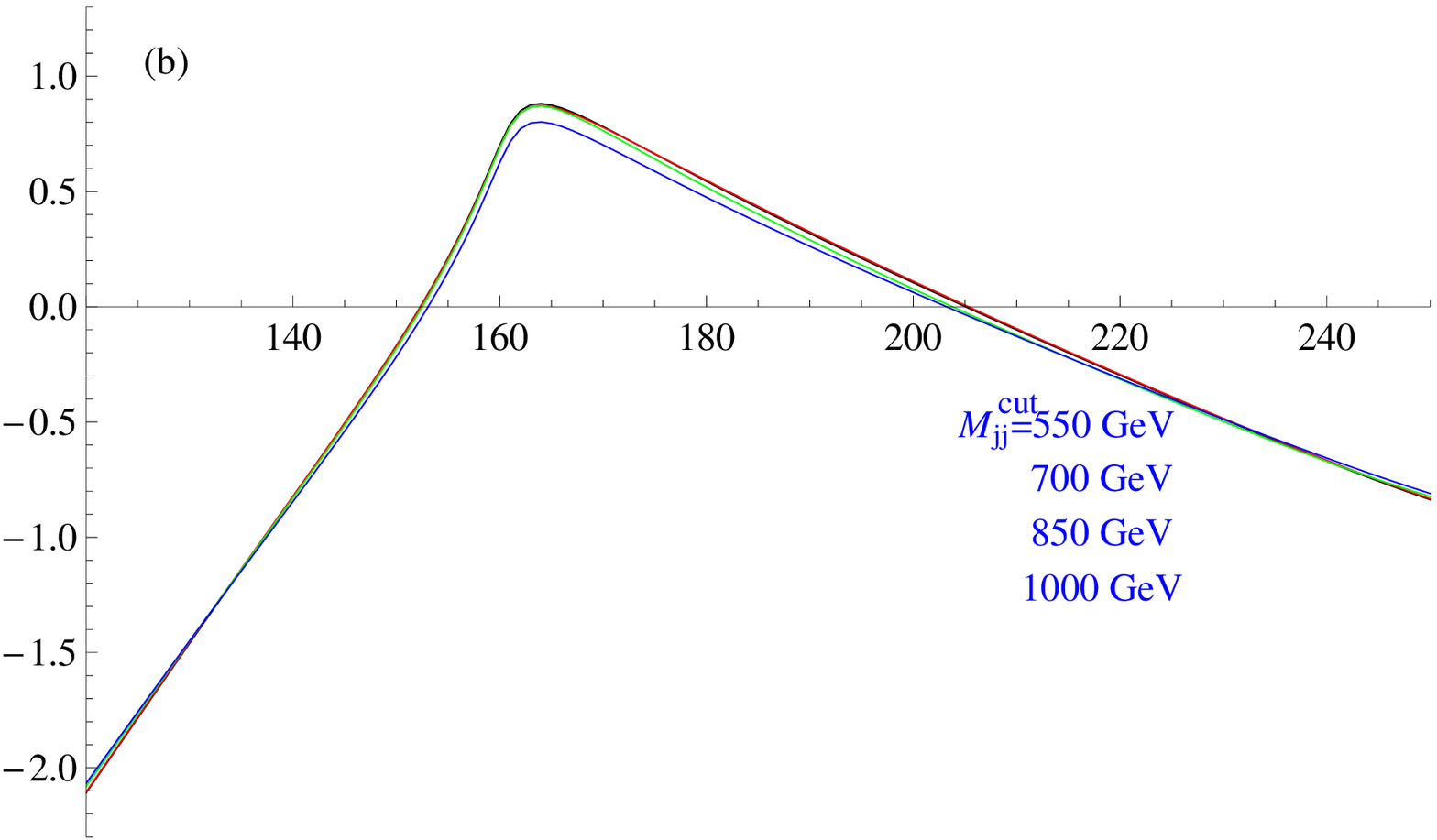}
\\
\includegraphics[width=8cm]{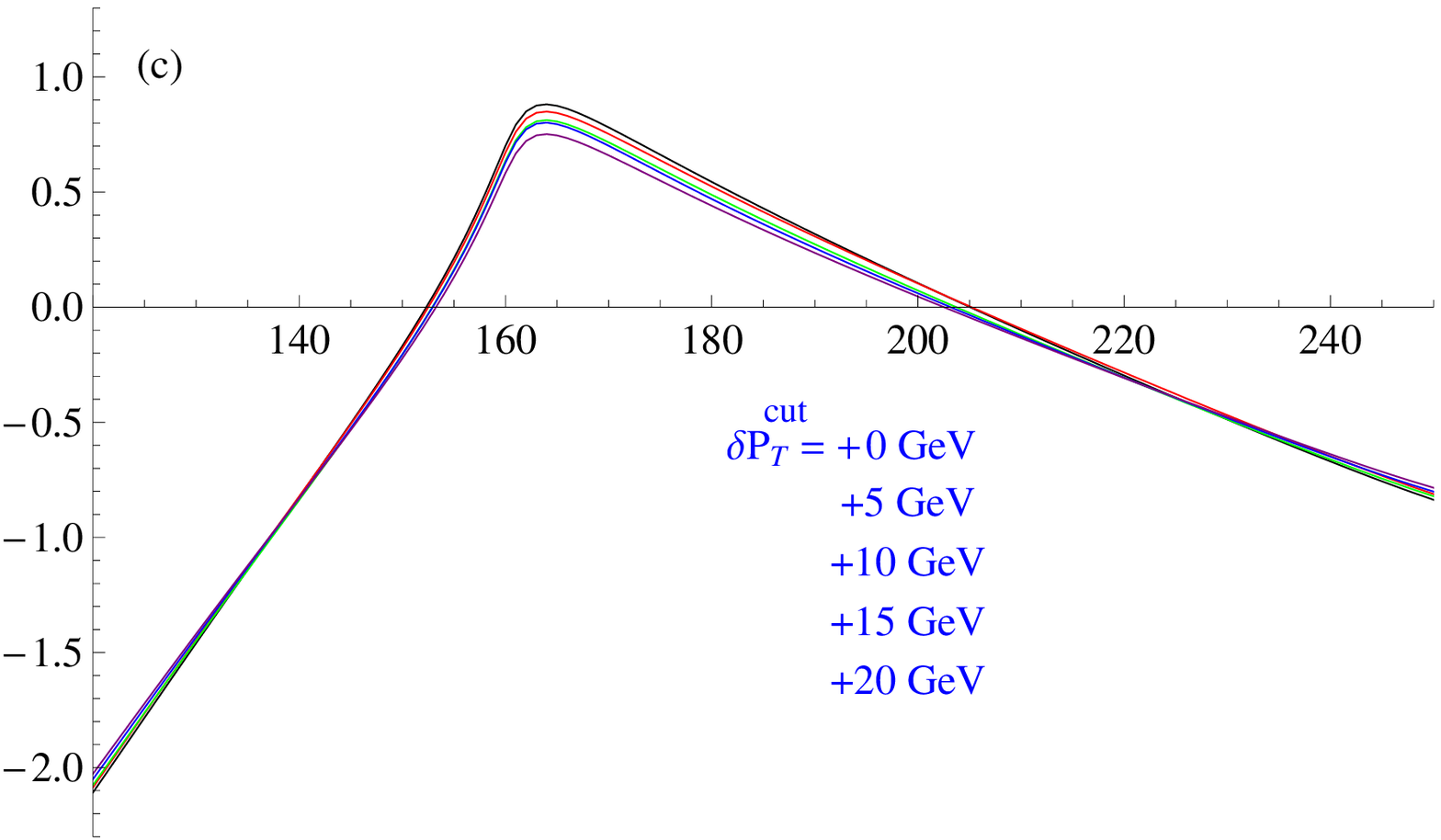}
\includegraphics[width=8cm]{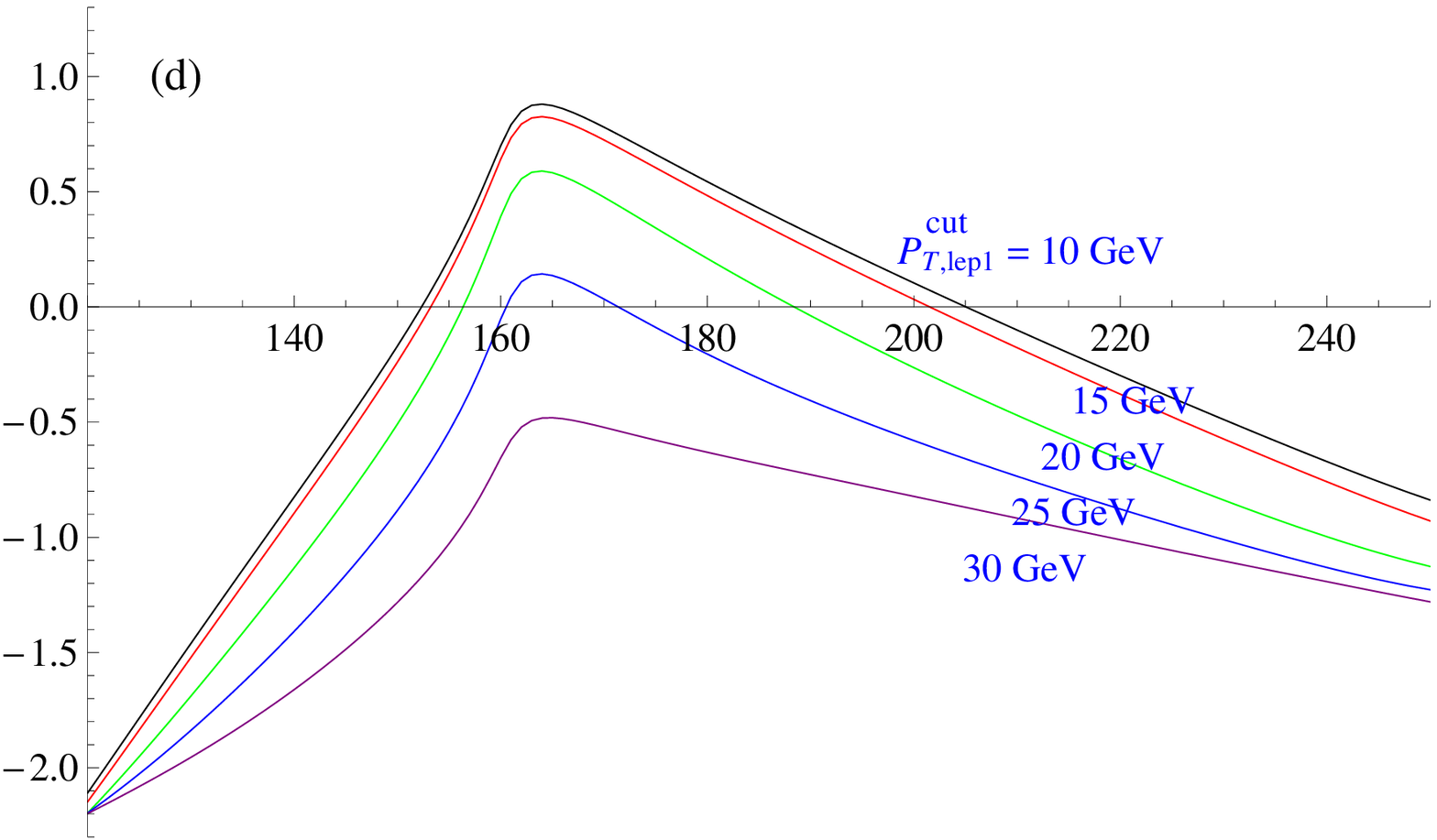}
\\
\includegraphics[width=8cm]{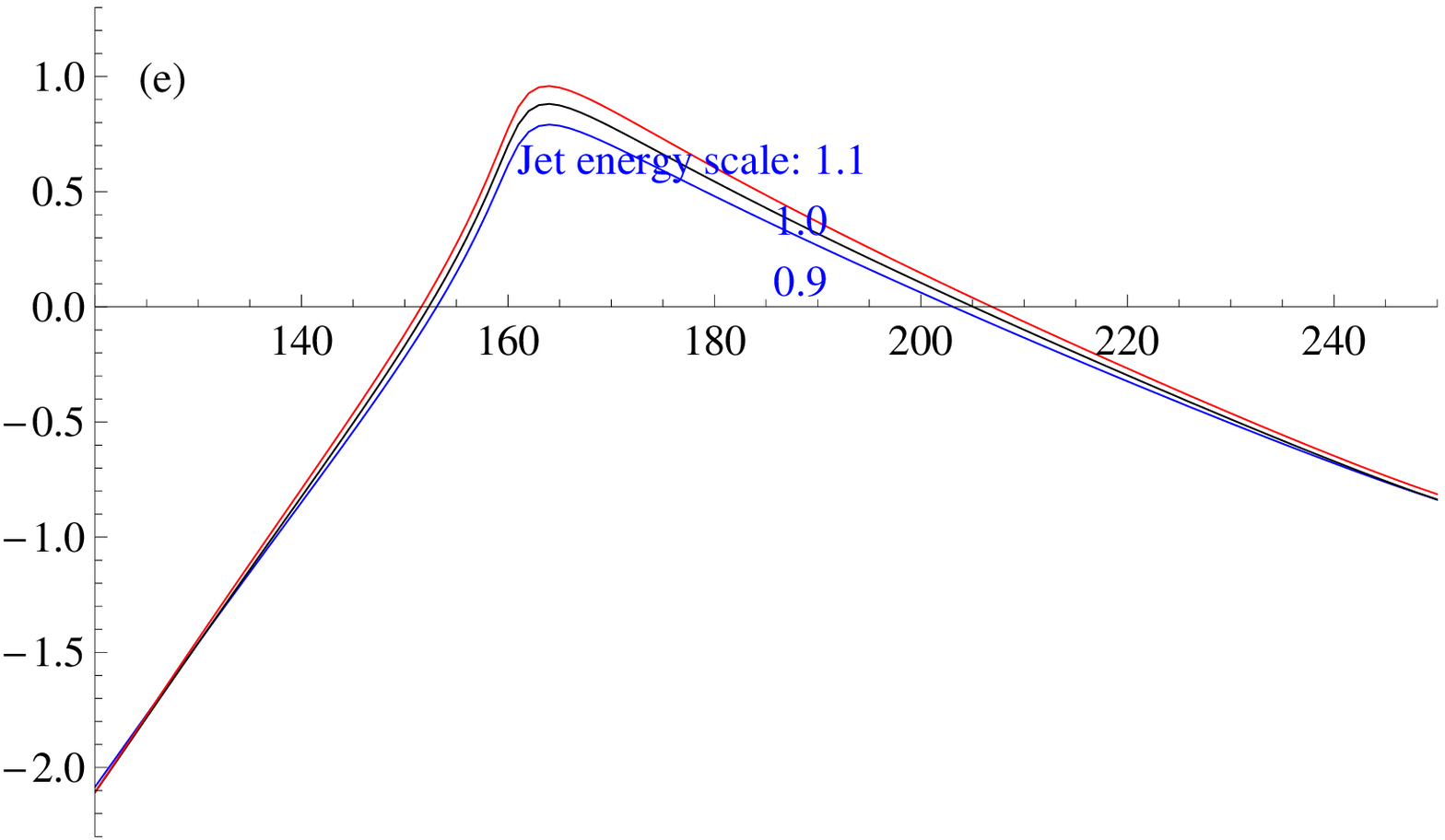}
\includegraphics[width=8cm]{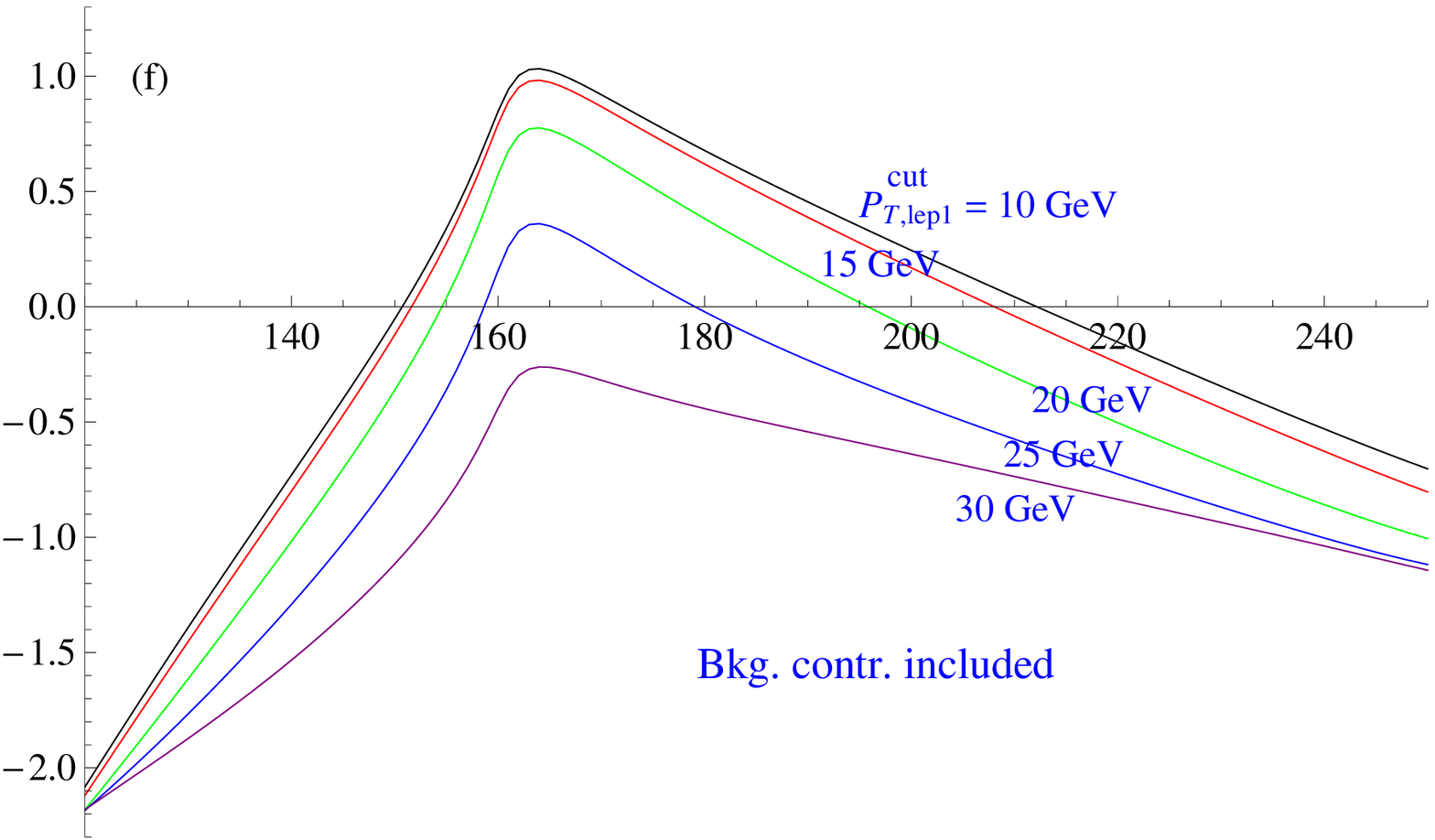}
\vspace*{3mm}
\end{tabular}
\caption{\label{Fig-VaryCuts}
$10^5\times\langle {\cal O}_G \rangle$ vs.\
$m_H$ (parameter in
$d\Gamma_{H\to \ell\ell\nu\nu}/dE_0|_{M_{\ell\ell}<45{\rm GeV}}$).
MC events with $m_H^{\rm MC}=150$~GeV and for $\mu\mu$ mode are used.
Through (a)--(e) only the signal events are used, while in (f)
both signal and background events are used.
All the cut values, except the ones shown explicitly in figs.~(b)(c)(d)(f),
are kept fixed to their default values.
}
\vspace*{-7mm}
\end{figure}
~
\end{widetext}

Now we examine stability and reliability of our prediction, using the $\mu\mu$
mode and MC input $m_H^{\rm MC}=150$~GeV.
In Fig.~\ref{Fig-VaryCuts}(a) we vary the factorization 
scale $\mu_{\rm fac}$ in PDF and PYTHIA and plot $\langle {\cal O}_G \rangle$
as a function of $m_H$:
$\mu_{\rm fac}$ is taken as $1/2$, $1$ and $2$ times its 
default value of MadEvent ($P_T$ of each scattered parton
from each proton in the vector-boson fusion case).\footnote{
Our MC event generator being leading order, 
there is a significant scale dependence, for instance, in the  
distribution of the number of jets.
}
$\delta m_H$ (and hence the value of $m_H$ where 
$\langle {\cal O}_G(m_H) \rangle$=0)
changes by
about 0.8~GeV as $\mu_{\rm fac}$ is varied from $1/2$ to 2.
In Fig.~\ref{Fig-VaryCuts}(b) we vary the cut value of
the invariant mass of tagged
two jets from 550~GeV to 1~TeV.
Corresponding variation of $\delta m_H$ is about 0.8~GeV.
In Fig.~\ref{Fig-VaryCuts}(c) we vary the values of the $P_T$ cuts
of tagged jets, $P_{T,j1}>40~{\rm GeV}+\delta P_{T,j}$ and  
$P_{T,j2}>20~{\rm GeV}+\delta P_{T,j}$,
between $\delta P_{T,j}=0$ and 20~GeV.
($\delta P_{T,j}=0$ is the default value.)
Variation of $\delta m_H$ is about 1.0~GeV.
These features agree with our expectation that $m_H$
determination is insensitive to PDF and cuts involving only jets,
since only the $\beta$ distribution of the 
Higgs boson would be affected.
Moreover, we find that $\langle {\cal O}_G \rangle$
as a function of $m_H$ is also fairly stable.

On the other hand, $\delta m_H$ and 
$\langle {\cal O}_G(m_H) \rangle$ depend strongly on the lepton cuts.
In Fig.~\ref{Fig-VaryCuts}(d) we vary the values of the $P_T$ cut
of the leading muon between 10~GeV and 30~GeV (while keeping the
$P_T$ cut value of the subleading muon to 10~GeV).
The line $\langle {\cal O}_G(m_H) \rangle$ moves downwards and
changes shape as the cut becomes tighter.
This stems from the fact that the $P_T$ cut suppresses lower part
of the lepton energy distribution.

Cuts involving missing transverse momentum
$\vec{P}_T^{\rm miss}$($=-\vec{P}_T^{\rm had}-\vec{P}_T^{\ell\ell}$)
may also affect the $m_H$ determination, since
the  cuts affect the lepton energy distribution through indirect restrictions
on $\vec{P}_T^{\ell\ell}$.
Here, $\vec{P}_T^{\ell\ell}$ denotes the transverse momentum of
the tagged dilepton system, while $\vec{P}_T^{\rm had}$ denotes the
sum of the transverse momenta of all other visible particles.
Since magnitudes of systematic uncertainties in
$\vec{P}_T^{\rm had}$ and $\vec{P}_T^{\ell\ell}$ measurements
are rather different,
instead of varying cuts involving $\vec{P}_T^{\rm miss}$,
we multiply $|\vec{P}_T^{\rm had}|$ by a scale factor of 0.9, 1.0 and 1.1;
we keep all the cut values unchanged.
See Fig.~\ref{Fig-VaryCuts}(e).
This variation of energy scale
affects restrictions on $\vec{P}_T^{\ell\ell}$ indirectly,
since bounds on 
$\vec{P}_T^{\rm miss}=-\vec{P}_T^{\rm had}-\vec{P}_T^{\ell\ell}$
are kept fixed.
We find that variation of $\delta m_H$ is about 1.6~GeV.

In Fig.~\ref{Fig-VaryCuts}(f) we include the background contributions
and vary the leading lepton $P_T$ cut.
(Through Fig.~\ref{Fig-VaryCuts}(a)--(e) only leptons from the signal events
are used.)
Compared to Fig.~\ref{Fig-VaryCuts}(d),
each line moves upwards and the line shape is modified slightly
by inclusion of the backgrounds. 
We have already seen this effect in the shift of the reconstructed
Higgs mass in Fig.~\ref{Fig4} and
Tab.~\ref{delmH}.

From these examinations and similar examinations for other
input Higgs mass values and decay modes, we estimate that our prediction
is fairly insensitive to uncertainties in PDF and jet variables.\footnote{
This feature persists with higher lepton $P_T$ cuts.
For example, with a cut
$P_{T,{\rm lep1}}\!>\! 25$\,GeV,
Figs.\,\ref{Fig-VaryCuts}(a)(b)(c)(e)
look qualitatively similar,  except that 
all the lines move downwards. 
%In particular, in each figure, variation among all the lines 
%in horizontal direction
%around $m_H= 150$\,GeV is almost the same as or less than the original.
}
On the other hand, the prediction is strongly dependent on the lepton $P_T$ cuts.
It is dependent also on other
lepton acceptance corrections.
These effects with respect to only leptons can in principle be estimated accurately,
by understanding detector coverage and detector performances well.

We note that all the lines in Fig.~\ref{Fig-VaryCuts}(f)
can be plotted using the real experimental data and can be compared with
the prediction.
Since these lines can be predicted accurately and
are dependent on MC input Higgs mass value $m_H^{\rm MC}$,
we can determine the Higgs mass by a fit of 
$\langle {\cal O}_G(m_H) \rangle$, 
provided the background contributions can be estimated accurately.
Similarly, all the lines in Figs.~\ref{Fig-VaryCuts}(b) and (c) can be compared,
after inclusion of background contributions, with the corresponding
ones plotted using the real experimental data.
Alternatively we may make similar comparisons by
loosening various cuts.
This procedure raises relative weight of background contributions,
so that we can test the prediction for the
background contributions.
We may also make use of the large degree of freedom
of the observable ${\cal O}_G$ for further tests
of the prediction.

In the case that higher lepton $P_T$ cuts are unavoidable,
determination of the Higgs mass by a fit of 
$\langle {\cal O}_G(m_H) \rangle$ would be more
realistic, as mentioned above, since 
$\langle {\cal O}_G(m_H) \rangle$ is shifted considerably.\footnote{
Recently requirement for
the single lepton trigger has become severer 
in the LHC experiments \cite{ATLASnote134}.
Hence, predictions with higher lepton $P_T$ cut
values may be more realistic.
In principle, we can also use the dilepton trigger, in which case
lower $P_T$ cut values may be used.
}
Explicitly, a fit can be carried out in the following way.
We can compute $\langle {\cal O}_G(m_H) \rangle$
using lepton energy distributions measured in a real experiment
and MC simulation, respectively, as
\bea
\langle {\cal O}_G(m_H) \rangle_{\rm exp}\equiv
\int dE_\ell\, D_{\rm exp}(E_\ell;m_H^{\rm true}){\cal O}_G(E_\ell;m_H),
&&
\\
\langle {\cal O}_G(m_H) \rangle_{\rm MC}\equiv
\int dE_\ell\, D_{\rm MC}(E_\ell;m_H^{\rm MC}){\cal O}_G(E_\ell;m_H).
&&
\eea
Defining a distance between $\langle {\cal O}_G(m_H) \rangle_{\rm exp}$
and $\langle {\cal O}_G(m_H) \rangle_{\rm MC}$ as
\bea
&&
d^2(m_H^{\rm MC})\equiv \int_{m_1}^{m_2}\!\!\!\!dm_H\,
w(m_H)\,
\nonumber\\
&&~~~~~~~~~~~~~~~
\times
[
\langle {\cal O}_G(m_H) \rangle_{\rm exp}
-
\langle {\cal O}_G(m_H) \rangle_{\rm MC}
]^2 \, ,
\eea
and minimizing $d(m_H^{\rm MC})$,
we may reconstruct the Higgs mass.
Here, $w(m_H)\geq 0$ denotes an appropriate weight.
As 
the lepton $P_T$ cuts become tighter, sensitivity to the Higgs mass 
decreases.
For instance, in the case $m_H^{\rm true}=150$~GeV,
if we impose 
$P_{T,{\rm lep1}}>25$~GeV and
$P_{T,{\rm lep2}}>10$~GeV to the leading and subleading leptons,
respectively,
we estimate that the statistical error of the Higgs mass,
reconstructed by the above fit,
to be about $5.0$~GeV for an integrated luminosity of
100~fb$^{-1}$.
The size of error is not very dependent on choices of
$m_1$ and $m_2$ or the weight $w$.
(We took $m_1=120$~GeV, $m_2=240$~GeV and $w=1$.)

Comparing Tab.~\ref{tabstat-err}
and the examination in Figs.~\ref{Fig-VaryCuts}, 
we anticipate that the statistical error
will dominate over systematic errors.
The accuracy of $m_H$ determination is better
for a lighter Higgs mass, within the range $150~{\rm GeV}\simlt m_H \simlt 200$~GeV. 
(e.g.\ If $m_H=150$~GeV, it would be reconstructed with 2--3\% accuracy
with 100~fb$^{-1}$ integrated luminosity.)
% Out of this mass range,
% statistics of signal events may be too small with the cuts
% used in this analysis; we may adjust cuts and improve 
% statistics in such cases.
%We relegate the study to our future work.

Let us comment on the effects of leptons from $\tau$'s, which
we have neglected.
Since these leptons would have a lower energy spectrum, the effects
of the lepton $P_T$ cuts are likely to be enhanced.
We anticipate from Fig.~\ref{Fig-VaryCuts}(d) that the effects
are to move the line $\langle {\cal O}_G(m_H) \rangle$ downwards.
Nevertheless, we expect that
the additional contribution from $\tau$'s and the cut effect
will be accurately predictable and should not deteriorate
the $m_H$ determination.

The main purpose of the present study on $m_H$ reconstruction
is to demonstrate that the mass reconstruction using ${\cal O}_G$ is
applicable to such a 
complicated process.
We chose the $WW$--fusion mode (rather than $gg$--fusion mode)
since a good signal-to-noise
ratio is beneficial in demonstrating clearly
the characteristics of the present method.
% Indeed we confirmed $|\delta m_H/m_H|\ll 1$, which shows that deviations
% from the ideal limit is suppressed and kept under control.
% It is a unique property of our method that we have a good
% handle on estimates of systematic
% errors through the value of $\delta m_H$.
We confirmed stability of our prediction except against 
purely leptonic cuts.
Good understanding of background contributions is also important.

Experimentally $m_H$ would 
be reconstructed very accurately using the decay modes
$H\to ZZ^{(*)}\to \ell\ell\ell\ell$ in the relevant mass range
\cite{Ball:2007zza}.
The $WW^{(*)}$ decay modes can be used to test consistency of the
Higgs mass values,
using our method or those of \cite{Davatz:2006ja,Barr:2009mx}.
Ref.~\cite{Davatz:2006ja} uses leptonic variables but 
depends severely on PDF.
We reemphasize that our method is independent of
PDF in the leading-order. 
In a future work, we may examine the $gg$--fusion mode,
in which the statistical error would be smaller while systematic
errors would be larger than the $WW$--fusion mode.

The observables proposed in this paper would have wide applications.
For example, in supersymmetric models, ${\cal O}_G$ can be applied to 
a slepton decay into
a lepton plus an undetected particle;
since it is a two-body decay, its analysis would be straightforward.
Another possible application is a measurement of the top quark mass.
A preliminary analysis indicates that the top quark polarization effects
are small.

\medbreak
The authors express their condolences for victims of disasters in Japan.
The disasters also struck the current project heavily.
The authors hope the completion of this work to be a step forward in revival
from the tragedy.
% H.Y.\ wishes to thank the members of 
% Tohoku Univ.\ Particle Theory Group for their
% warm hospitality.
The work of Sumino is supported in part by Grant-in-Aid for
scientific research No.\ 23540281 from
MEXT, Japan.


\end{document}